\documentclass{llncs}

\usepackage{latexsym}
\usepackage{amsmath}
\usepackage{amssymb}
\usepackage{mathpartir}
\usepackage{semantic}
\usepackage{stmaryrd}
\usepackage{bussproofs}
\usepackage[usenames]{color}
\usepackage[latin1]{inputenc} 
\usepackage{graphicx}
\usepackage{url}

\newcommand{\cminor}{Cminor}
\newcommand{\compcert}{CompCert}

\newcommand{\defeq}{=_{\mathrm{def}}}

\newcommand{\infrencee}[2]{
\inference{\hspace*{-8pt}#1\hspace*{-8pt}}{\hspace*{-8pt} #2\hspace*{-8pt}}}

\newcommand{\guardbox}{\raisebox{1pt}{\makebox[0pt][l]{\(\sqcap\)}}{\raisebox{-1pt}{\(\sqcup\)}}}

\newcommand{\tyface}[1]{\ensuremath{\mathsf{#1}}}
\newcommand{\option}{\tyface{option}}
\newcommand{\val}{\tyface{val}}
\newcommand{\Vundef}{\tyface{Vundef}}
\newcommand{\Vint}{\tyface{Vint}}
\newcommand{\Vfloat}{\tyface{Vfloat}}
\newcommand{\Vptr}{\tyface{Vptr}}
\newcommand{\expr}{\tyface{expr}}
\newcommand{\env}{\tyface{env}}
\newcommand{\mem}{\tyface{mem}}
\newcommand{\exprlist}{\tyface{exprlist}}

\newcommand{\Eval}{\tyface{Eval}}
\newcommand{\Evar}{\tyface{Evar}}
\newcommand{\Eop}{\tyface{Eop}}
\newcommand{\Eload}{\tyface{Eload}}

\newcommand{\Enil}{\tyface{Enil}}
\newcommand{\Econs}{\tyface{Econs}}
\newcommand{\Sassign}[2]{#1:=#2}
\newcommand{\Sstore}[3]{\tyface{[}#2\tyface{]}_{#1}\tyface{:=}#3}
\newcommand{\Scall}[4]{\tyface{call}\,#1\,#3\,#4}
\newcommand{\Sif}[3]{\tyface{if}\,#1\,\tyface{then}\,#2\,\tyface{else}\,#3}
\newcommand{\Sloop}[1]{\tyface{loop}\,#1}
\newcommand{\Sblock}[1]{\tyface{block}\,#1}
\newcommand{\Sexit}[1]{\tyface{exit}\,#1}
\newcommand{\Sreturn}[1]{\tyface{return}\,#1}
\newcommand{\Sseq}[2]{#1;#2}
\newcommand{\Sskip}{\tyface{skip}}

\newcommand{\Kstop}{\tyface{Kstop}}
\newcommand{\Kseq}[2]{#1\cdot #2}
\newcommand{\Kblock}[1]{\tyface{Kblock\,#1}}
\newcommand{\Kcall}[5]{\tyface{Kcall}\,#1\,#2\,#3\,#4\,#5}

\newcommand{\exitcont}[2]{\Kseq{\Sexit{#1}}{#2}}
\newcommand{\returncont}[2]{\Kseq{\Sreturn{#1}}{#2}}

\newcommand{\guard}[3]{#2\: {\guardbox{}_{#1}}\: #3}

\newcommand{\rguard}[3]{#2\: \makebox[0pt][l]{\scriptsize \hspace{1pt}\textsc{r}}{\guardbox{}_{#1}}\: #3}

\newcommand{\bguard}[3]{#2\: \makebox[0pt][l]{\scriptsize \hspace{1pt}\textsc{b}}{\guardbox{}_{#1}}\: #3}

\newcommand{\Cont}[2]{(#1,#2)}

\newcommand{\st}{\sigma}

\newcommand{\ctl}{\kappa} 
\newcommand{\stmt}{\tyface{stmt}}

\newcommand{\continuation}{\tyface{continuation}}
\newcommand{\id}{\ensuremath{\mathit{id}}}
\newcommand{\fmap}{\Psi}

\newcommand{\infootprint}[4]{#4 \vdash \tyface{#3}_#2\, #1}
\newcommand{\notinfootprint}[4]{#4 \not\vdash \tyface{#3}_#2\, #1}
\newcommand{\evalexpr}[7]{#1;(#2;#3;#4;#5) \vdash #6 \Downarrow #7}
\newcommand{\evalexp}[3]{\fmap; #1 \vdash #2 \Downarrow #3}
\newcommand{\evaloperation}[5]{#1;#2 \vdash #3(#4) \Downarrow_\tyface{eval\_operation} #5}

\newcommand{\loadv}[4]{#1\vdash #2\stackrel{#3}{\mapsto}#4}
\newcommand{\storev}[5]{#5=#2[#3\stackrel{#1}{:=}#4]}

\newcommand{\step}[2]{\fmap \vdash #1\:\longmapsto\:#2}

\newcommand{\stepn}[3]{\fmap \vdash #1\:{\longmapsto^{#3}}\:#2}

\newcommand{\safe}[1]{\tyface{safe}\,#1}
\newcommand{\absorb}[3]{\tyface{absorb}(#1,#2,#3)}
\newcommand{\cat}[2]{#1 \circ #2}

\newcommand{\semax}[6]{#1;#2;#3\vdash \{#4\}#5\{#6\}}
\newcommand{\emp}{\ensuremath{\mathrm{\mathbf{emp}}}}
\newcommand{\TT}{\ensuremath{\mathrm{\mathbf{true}}}}
\newcommand{\FF}{\ensuremath{\mathrm{\mathbf{false}}}}
\newcommand{\pure}[1]{\tyface{pure}\,#1}
\newcommand{\assertprop}[1]{\lceil #1 \rceil}
\newcommand{\assertexpr}[1]{\lceil #1 \rceil_\tyface{expr}}
\newcommand{\assertmapsto}[4][]{#2\stackrel{#3}{\mapsto}_{#1}#4}

\newcommand{\assertdef}[1]{\mathsf{defined}(#1)}

\newcommand{\ceq}[3]{#2\stackrel{#1}{\mbox{\texttt{==}}}#3}
\newcommand{\ieq}[2]{\ceq{\mathrm{int}}{#1}{#2}}

\newcommand{\Bnil}{\tyface{nil}_\tyface{B}}
\newcommand{\Bcons}[2]{#1\cdot #2}

\DeclareMathOperator{\dom}{dom}

\begin{document}

\author{
Andrew~W.~Appel\inst{1}\fnmsep{}\thanks{Appel supported in part by NSF Grants CCF-0540914 and CNS-0627650.
This work was done, in part, while both authors were on sabbatical at INRIA.}
  \and
Sandrine Blazy\inst{2}\fnmsep{}$^\star$
}
\institute{Princeton University \and ENSIIE}

\authorrunning{Appel \& Blazy}
\title{Separation Logic for Small-step \cminor}

\maketitle 

\pagestyle{plain}

\vspace{-20pt}
\begin{abstract}
\cminor{} is a mid-level imperative programming language;
there are proved-correct optimizing compilers
from C to \cminor{} and from \cminor{} to machine language.
We have redesigned \cminor{} so that it is suitable for Hoare Logic
reasoning and we have designed a Separation Logic for \cminor.
In this paper, we give a small-step semantics (instead of the big-step of
the proved-correct compiler) 
that is motivated by the need to support future concurrent extensions.
We detail a machine-checked proof of soundness of our Separation Logic.
This is the first large-scale machine-checked 
proof of a Separation Logic w.r.t. a small-step semantics.
The work presented in this paper has been carried out in the Coq proof assistant. 
It is a first step towards an
environment in which concurrent \cminor{} programs 
can be verified using Separation Logic
and also
compiled by a proved-correct compiler with formal
end-to-end correctness guarantees.
\end{abstract}

\section{Introduction}

The future of program
verification is to connect machine-verified source programs to
machine-verified compilers, and run the object code on 
machine-verified hardware.  To connect the verifications end to end,
the source language should be specified as a structural operational semantics
(SOS) represented
in a logical framework; the target architecture can also be specified
that way.  Proofs of source code can be done in the logical framework,
or by other tools whose soundness is proved w.r.t. the SOS
specification; 
these may be in safety proofs via type-checking,
correctness proofs via Hoare Logic,
or (in source languages designed for the purpose) 
correctness proofs by a more expressive proof theory.
The compiler---if it
is an optimizing compiler---will be a stack of
phases, each with a well specified SOS of its own.  There
will be proofs of (partial) correctness of each compiler phase, or
witness-driven recognizers for correct compilations, w.r.t.
the SOS's that are inputs and outputs to the phases.

Machine-verified hardware/compiler/application stacks have been built
before.  Moore described a verified compiler
for a ``high-level assembly language'' \cite{moore89}.
Leinenbach \emph{et al.} \cite{leinenbach05} have built
and proved a compiler for \emph{C0}, a small C-like language, 
as part of a project to build machine-checked
correctness proofs 
of source programs, Hoare Logic, compiler, micro-kernel, and RISC processor.
These are both simple one- or two-pass
nonoptimizing compilers.

Leroy \cite{leroy06} has built and proved correct in Coq~\cite{COQ-www}
a compiler called \emph{CompCert} from a high-level intermediate language \emph{\cminor}
to assembly language for the Power PC architecture.
This compiler has 4 intermediate languages, allowing
optimizations at several natural levels of abstraction.
Blazy \emph{et al.} have built and proved correct
a translator from a subset of C to \cminor{} \cite{blazy06:fm}.
Another compiler phase on top (not yet implemented)
will then yield a
proved-correct compiler from C to machine language.
We should therefore reevaluate the conventional wisdom that
an entire practical optimizing compiler cannot be proved correct.

A software system can have
components written in different languages, and we would like
end-to-end correctness proofs of the whole system.  
For this, we propose a new variant of \cminor{} as a machine-independent intermediate language to serve
as a common denominator between high-level languages.
Our new \cminor{} has a usable Hoare Logic, so that 
correctness proofs for some components can be done directly at the level of \cminor.

\cminor{} has a
``calculus-like'' view of local variables and procedures
(\textit{i.e.} local variables are bound in an environment), 
while Leinenbach's C0 has a ``storage-allocation'' view (\textit{i.e.} local variables are stored in the stack frame).  
The calculus-like view
will lead to easier reasoning about program transformations
and easier use of \cminor{} as a target language, and fits naturally
with a multi-pass optimizing compiler such as \compcert;
the storage-allocation view suits the 
one-pass nonoptimizing C0 compiler and can accommodate in-line assembly code.

\cminor{} is a promising candidate
as a common intermediate language for end-to-end correctness proofs.
But we have many demands on our new variant of \cminor, 
only the first three of which are 
satisfied by Leroy's \cminor.
\begin{itemize}
\item[$\bullet$] \cminor{} has an operational semantics
represented in a logical framework.
\item[$\bullet$] There is a proved-correct compiler from \cminor{} to machine language.
\item[$\bullet$] \cminor{} is usable as the high-level target language of
a C compiler.
\item[$\circ$] Our semantics is a \emph{small-step} semantics, 
to support reasoning about input/output, concurrency,
and nontermination.
\item[$\circ$] \cminor{} is machine-independent 
over machines in the ``standard model''
(\textit{i.e.} 32- or 64-bit single-address-space byte-addressable multiprocessors).
\item[$\circ$] \cminor{} can be used as a mid-level target language of an ML compiler \cite{dargaye:jfla07},
or of an OO-language compiler, so that we can
integrate correctness proofs of ML or OO programs with the
proofs of their run-time systems and libraries.
\item[$\circ$] As we show in this paper, \cminor{} supports an axiomatic Hoare Logic 
(in fact, Separation Logic), proved sound
with respect to the small-step semantics, for reasoning about low-level 
(C-like) programs.
\item[$\circ$] 
In future work, we plan to extend \cminor{} to
\label{L-Concur}
 be concurrent in  the ``standard model'' of thread-based 
preemptive lock-synchronized weakly consistent shared-memory programming.
The sequential soundness proofs we present here should be reusable
in a concurrent setting, as we will explain.
\end{itemize}

Leroy's original \cminor{} had several Power-PC dependencies,
is slightly clumsy to use as the target of an ML compiler,
and is a bit clumsy to use in Hoare-style reasoning.
But most important, Leroy's semantics is a big-step semantics
that can be used only to reason about terminating sequential 
programs.
We have redesigned \cminor's syntax and semantics 
to achieve all of these goals.
That part of the redesign to achieve target-machine portability was
done by Leroy himself.  
Our redesign to ease its use as an ML back end
and for Hoare Logic reasoning was fairly simple.
Henceforth in this paper, \cminor{} will refer to the new version of the \cminor{} language.

The main contributions of this paper are
a small-step semantics suitable for
compilation and for Hoare Logic; and
the first machine-checked proof of soundness of 
a sequential Hoare Logic (Separation Logic) w.r.t.
a small-step semantics.
Schirmer \cite{schirmer06:phd} has a machine-checked \emph{big-step}
Hoare-Logic soundness proof for a control flow much like ours,
extended by Klein \emph{et al.}  \cite{Klein:POPL07} to a
C-like memory model.
Ni and Shao \cite{Shao:popl06} have a machine-checked proof of soundness of a Hoare-like logic w.r.t.
a small-step semantics, but for an assembly language and for much simpler assertions
than ours.

\section{Big-step Expression Semantics}\label{big-step}

The C standard \cite{ansi88c} describes a memory model
that is byte- and word-addressable
(yet portable to big-endian and little-endian machines)
with a nontrivial semantics
for uninitialized variables.
Blazy and Leroy formalized this model
\cite{blazy05:icfem} for the semantics of \cminor.
In C, pointer
arithmetic within any 
malloc'ed block is defined, but pointer arithmetic between different
blocks is undefined; \cminor{} therefore has non-null pointer values
comprising an abstract block-number and an \tyface{int} offset.
A \texttt{NULL} pointer is represented by the integer value $0$.
Pointer arithmetic between blocks, and
reading uninitialized variables, are undefined but not 
illegal: expressions in \cminor{} can evaluate to
\emph{undefined} (\tyface{Vundef}) without getting stuck.  

Each memory load or store is to a non-null
pointer value 
with a ``chunk'' descriptor
$\mathit{ch}$
specifying number of bytes,
signed or unsigned, int or float.
Storing as 32-bit-int then loading as 8-bit-signed-byte
leads to an undefined value.
Load and store operations on memory,
$\loadv{m}{v_1}{\mathit{ch}}{v_2}$ and 
$\storev{\mathit{ch}}{m}{v_1}{v_2}{m'}$,
are partial functions that yield results 
only if reading (resp., writing)
a chunk of type $\mathit{ch}$ at address $v_1$ is legal.
We write $\loadv{m}{v_1}{\mathit{ch}}{v}$ to mean
that the result of loading from memory $m$ at address $v_1$
a chunk-type $\mathit{ch}$ is the value $v$.

The \emph{values} of \cminor{} are \emph{undefined} (\tyface{Vundef}),
integers, pointers, and floats.  The \tyface{int} type
is an abstract data-type of 32-bit modular arithmetic.
The expressions of \cminor{} are literals, variables,
primitive operators applied to arguments,
and memory loads. 

There are 33 primitive \tyface{operation} symbols $\mathit{op}$;
two of these are for accessing global names and
local stack-blocks, and the rest is for integer and
floating-point arithmetic and comparisons.
Among these operation symbols are casts.
\cminor{} casts correspond to all portable C casts.
\cminor{} has an infinite supply \tyface{ident} of
variable and function identifiers $\mathit{id}$.  
As in C, there are two namespaces---each 
$\mathit{id}$ can be interpreted in a local
scope (using $\Evar\,(\mathit{id})$) or in a global scope
(using 
$\Eop$ with the operation symbol for accessing global names).

\vspace{-20pt}
\begin{align*}
i~:\tyface{int} ~ ::= ~ & [0,2^{32}) \\
 v~:~\val ~ ::= ~ & \Vundef{} ~|~ \Vint{}~(i)
~|~ \Vptr{}~(b,i) ~|~ \Vfloat{} ~(f)\\
e:\expr ~  ~ ::= & ~ \Eval~(v) ~|~
\Evar~(\id{}) ~|~ \Eop~(\mathit{op}, \mathit{el})~|~ 
\Eload~(\mathit{ch},e )\\
 \mathit{el}:\exprlist{} ~ ::= & ~\Enil ~|~ \Econs~(e,\mathit{el})
\end{align*}

\paragraph{Expression Evaluation.}
In original \cminor{}, expression evaluation is expressed by an inductive big-step relation.
Big-step statement execution is problematic for concurrency, but
big-step \emph{expression} evaluation is fine even for concurrent
programs, since we will use the separation logic to prove
noninterference.

Evaluation is deterministic.
Leroy chose to represent evaluation as a relation because 
Coq had better
support for proof induction over relations than over function definitions.
We have chosen to represent
evaluation as a partial function; this makes some proofs easier
in some ways: $f(x)=f(x)$ is simpler than $f\,x\,y\Rightarrow f\,x\,z
\Rightarrow y=z$.
Before Coq's new functional induction tactic was available, 
we developed special-purpose tactics to enable these proofs.
Although we specify expression evaluation as a function in Coq,
we present evaluation as a judgment relation in Fig.~\ref{evalexpr-fig}.
Our evaluation function is
(proved) equivalent to the inductively defined 
judgment $~\evalexpr{\fmap}{\mathit{sp}}{\rho}{\phi}{m}{e}{v}~$
where:
\begin{description}
\vspace{-7pt}
\item{$\fmap$} is the ``program,'' consisting of a global environment
$(\tyface{ident}\rightarrow \option\,\tyface{block})$
mapping identifiers to function-pointers and other global constants,
and a global mapping $(\tyface{block}\rightarrow \option\,\tyface{function})$
that maps certain (``text-segment'') addresses to function definitions.
\item{$\mathit{sp}:~\tyface{block}.$}
The ``stack pointer'' giving the address and size of the memory block for 
stack-allocated local data in the current activation record.
\item{$\rho:~\env{}.$}
The local environment, a finite mapping
from identifiers to values.
\item{$\phi:~\tyface{footprint}.$}
It represents the memory used by the evaluation of an expression (or a statement).
It is a mapping from  memory addresses to permissions.  Leroy's \cminor{} has no 
footprints.
\item{$m:~\mem.$}
The memory, a finite mapping from blocks
to block contents \cite{blazy05:icfem}.  
Each block represents
the result of a C \texttt{malloc},
or a stack frame, a global static variable, or a function
code-pointer. 
A block content consists of the dimensions of the block (low and high bounds) 
plus a mapping from byte offsets to byte-sized memory cells.
\item{$e:~\expr.$}
The expression being evaluated. 
\item{$v:~\val.$} The value of the expression.
\end{description}

\begin{figure}
\vspace{-35pt}
\begin{mathpar}
\evalexpr{\fmap}{\mathit{sp}}{\rho}{\phi}{m}{\,\Eval\, (v)\,}{v}
\and
\inference{x\in \dom \rho}{
\evalexpr{\fmap}{\mathit{sp}}{\rho}{\phi}{m}{\,\Evar\, (x)\,}{\rho(x)}}
\and
\inference{
\evalexpr{\fmap}{\mathit{sp}}{\rho}{\phi}{m}{\mathit{el}}{\mathit{vl}}
\and
\evaloperation{\fmap}{\mathit{sp}}{\mathit{op}}{\mathit{vl}}{v}
}{
\evalexpr{\fmap}{\mathit{sp}}{\rho}{\phi}{m}{\,\Eop\,(\mathit{op},\mathit{el})\,}{v}}
\and
\inference{
\evalexpr{\fmap}{\mathit{sp}}{\rho}{\phi}{m}{e_1}{v_1} \and
\and \infootprint{v_1}{\mathit{ch}}{load}\phi
\and \loadv{m}{v_1}{\mathit{ch}}{v}
}
{
\evalexpr{\fmap}{\mathit{sp}}{\rho}{\phi}{m}{\,\Eload\,(\mathit{ch},e_1)\,}{v}}
\end{mathpar}
\vspace{-10pt}
\caption{Expression evaluation rules}
\label{evalexpr-fig}
\vspace{-10pt}
\end{figure}

Loads outside
the footprint will cause expression evaluation to get stuck.
Since the footprint may have different permissions for
loads than for stores to some addresses, we write 
$\infootprint{v}{\mathit{ch}}{load}\phi$
(or $\infootprint{v}{\mathit{ch}}{store}\phi$)
to mean that all the addresses from $v$ to
$v+\left|\mathit{ch}\right| -1$ are 
readable (or writable).

To model the possibility of exclusive read/write access or
shared read-only access, we write $\phi_0\oplus\phi_1=\phi$
for the ``disjoint'' sum of two footprints, 
where $\oplus$ is an associative and commutative 
operator with several properties such as
$\infootprint{v}{\mathit{ch}}{store}{\phi_0} \,\Rightarrow\,
\notinfootprint{v}{\mathit{ch}}{load}{\phi_1}$,
$\infootprint{v}{\mathit{ch}}{load}{\phi_0} \,\Rightarrow
\infootprint{v}{\mathit{ch}}{load}{\phi}$
and
$\infootprint{v}{\mathit{ch}}{store}{\phi_0} \,\Rightarrow
\infootprint{v}{\mathit{ch}}{store}{\phi}$.
One can think of $\phi$ as a set of fractional permissions
\cite{bornat05:popl},
with 0 meaning no permission, $0<x<1$ permitting read,
and 1 giving read/write permission.
A \tyface{store} permission can be split into
two or more \tyface{load} permissions,
which can be reconstituted to obtain a \tyface{store} permission.
Instead of fractions, we use a more general and powerful model of sharable 
permissions similar to one described by 
Parkinson \cite[Ch.~5]{parkinson05:phd}.

Most previous models of Separation Logic (\textit{e.g.}, Ishtiaq and O'Hearn
\cite{ishtiaq01}) represent heaps as partial functions
that can be combined with an operator like $\oplus$.  Of course,
a partial function can be represented as a pair of a domain set and
a total function.  Similarly, we represent heaps as a footprint plus a 
\cminor{} memory; this does not add any particular difficulty to the soundness proofs for
our Separation Logic.

To perform arithmetic and other operations,
in the third rule of Fig.~\ref{evalexpr-fig}, the judgment
$\evaloperation{\fmap}{\mathit{sp}}{\mathit{op}}{\mathit{vl}}{v}$
takes an operator $\mathit{op}$ applied to a list of values $\mathit{vl}$
and (if $\mathit{vl}$ contains appropriate values)
produces some value $v$.  
Operators 
that access global names and local stack-blocks
make use of $\fmap$ and $\mathit{sp}$ respectively to
return the address of a global name or a local stack-block 
address.

\paragraph{States.}
We shall bundle together $(\mathit{sp};\rho;\phi;m)$
and call it the \emph{state}, written as $\sigma$.
%
We write $\evalexp{\sigma}{e}{v}$ to mean
$\evalexpr{\fmap}{\mathit{sp}_\sigma}{\rho_\sigma}{\phi_\sigma}{m_\sigma}{e}{v}$.

\paragraph{Notation.}
We write $\sigma[:=\rho']$ to mean the state $\sigma$
with its environment component $\rho$ replaced by $\rho'$, and so on
(\textit{e.g.} see rules 2 and 3 of Fig.~\ref{fig-step} in Section~\ref{seplog}).


\paragraph{Fact.} 
$\evaloperation{\fmap}{\mathit{sp}}{\mathit{op}}{\mathit{vl}}{v}$
and $\loadv{m}{v_1}{\mathit{ch}}{v}$ are both deterministic
relations, \textit{i.e.} functions.

\begin{lemma}
$\evalexp{\sigma}{e}{v}$
is a deterministic relation. (Trivial by inspection.)
\end{lemma}

\begin{lemma}
For any value $v$, there is an expression $e$ such that
$\forall\sigma.~(\evalexp{\sigma}{e}{v})$.
\vspace{-3pt}
\begin{proof} 
Obvious; $e$ is simply $\Eval\,v$.
But it is important nonetheless: reasoning about programs by rewriting
and by Hoare Logic often requires this property,
and it was absent from Leroy's \cminor{} for
\Vundef{} and \Vptr{} values.
$\blacksquare$
\end{proof}\end{lemma}


An expression may fetch from several different memory
locations, or from the same location several times.  
Because $\Downarrow$ is deterministic, we cannot
model a situation where the memory is updated by another thread
after the first fetch and before the second.
But we want a semantics that describes real
executions on real machines.  The solution is to evaluate expressions
in a setting where we can guarantee \emph{noninterference}.
We will do this (in our extension to Concurrent \cminor)
by guaranteeing that the footprints $\phi$ of different
threads are disjoint.


\paragraph{Erased Expression Evaluation.}
The \cminor{} compiler (CompCert) is proved correct w.r.t. an operational semantics that does not use
footprints.
Any program that successfully evaluates with footprints will also evaluate ignoring footprints.
Thus, for sequential programs where we do not need noninterference,
it is sound to prove properties in a footprint semantics and compile in an erased semantics.  We formalize and prove this in the full technical report \cite{appel07:tr}.

\section{Small-step Statement Semantics}
The statements of sequential \cminor{} are:
\begin{align*}
s:\stmt ~  ~ ::= & ~ 
   \Sassign{x}{e} 
~|~ \Sstore{\mathit{ch}}{e_1}{e_2}
~ |~ \Sloop{s}
~|~ \Sblock{s}
~|~ \Sexit{n} \\
&|~ \Scall{\mathit{xl}}{\Sigma}{e}{\mathit{el}}
~|~ \Sreturn{\mathit{el}} 
|~ \Sseq{s_1}{s_2}
~|~ \Sif{e}{s_1}{s_2}
~|~ \Sskip. 
\end{align*}
The assignment $\Sassign{x}{e}$ puts the value of $e$ into the local variable $x$.
The store $\Sstore{\mathit{ch}}{e_1}{e_2}$ puts (the value of) $e_2$ into the
memory-chunk $\mathit{ch}$ at address given by (the value of) $e_1$.
(Local variables are not addressable; global variables and heap locations
are memory addresses.)
To model exits from nested loops,
$\Sblock{s}$ runs $s$, which should not terminate normally but
which should $\Sexit{n}$ from the $(n+1)^{th}$ enclosing block, and
$\Sloop{s}$ repeats $s$ infinitely or until it returns or exits.
$\Scall{\mathit{xl}}{\Sigma}{e}{\mathit{el}}$ 
calls function $e$ with parameters (by value) $\mathit{el}$
and results returned back into the variables $\mathit{xl}$.
$\Sreturn{\mathit{el}}$ evaluates and returns a sequence of results,
$(\Sseq{s_1}{s_2})$ executes $s_1$ followed by $s_2$ (unless 
$s_1$ returns or exits), and the statements
$\tyface{if}$ and $\Sskip$ are as the reader might expect.

Combined with infinite loops and $\tyface{if}$ statements, blocks and exits
suffice to express efficiently all reducible control-flow graphs, notably those arising from
C loops.
The C statements $\tyface{break}$ and $\tyface{continue}$ are translated as
appropriate $\tyface{exit}$ statements. Blazy \emph{et al.} \cite{blazy06:fm} detail the translation of these C statements
into \cminor.

\paragraph{Function Definitions.}
A program $\fmap$ comprises two mappings:
a mapping from function names to memory blocks (\textit{i.e.}, abstract
addresses),
and a mapping from memory blocks to function definitions.
Each function definition may be written as
$f=( 
\mathit{xl},\mathit{yl},n,s)$, where 
$\tyface{params}(f)=\mathit{xl}$ is a list of formal parameters,
$\tyface{locals}(f)=\mathit{yl}$ is a list of local variables,
$\tyface{stackspace}(f)=n$ is the size of the local stack-block to which
$\mathit{sp}$ points, and the statement $\tyface{body}(f)=s$ is the 
function body.  

\paragraph{Operational Semantics.}
Our small-step semantics for statements is based on continuations, 
mainly to allow a uniform representation of statement execution
that facilitates the design of lemmas.
Such a semantics also avoids all search rules (congruence rules),
which avoids induction over search rules in both
the Hoare-Logic soundness proof and the compiler correctness proof.\footnote{
We have proved in Coq the
equivalence of this small-step semantics with the 
big-step semantics of \compcert{} 
(for programs that terminate).} 

\begin{definition}
A \tyface{continuation} $k$ has
a state $\st$
and a \tyface{control} stack $\ctl$. There are 
sequential control operators
to handle local control flow ($\tyface{Kseq}$, written as
 $\cdot$), 
intraprocedural control flow $(\tyface{Kblock})$,
and function-return $(\tyface{Kcall})$; this last carries not only
a control aspect but an activation record of its own.
The control operator $\Kstop$ represents the safe
termination of the computation.
\vspace{-8pt}
\begin{align*}
\ctl:\tyface{control}~::= &~ \Kstop ~|~  
\Kseq s \ctl
~|~ \Kblock\ctl 
~|~ \Kcall{\mathit{xl}}{f}{\mathit{sp}}{\rho}\ctl \\
k:\continuation ::= &~\Cont{\st}{\ctl}
\end{align*}
\end{definition}

The sequential small-step function takes the
form $\step{k}{k'}$ (see Fig.~\ref{fig-step}),
and we define as usual its 
reflexive transitive closure $\longmapsto^{*}$.
As in C, there is no boolean type in \cminor.
In Fig.~\ref{fig-step}, the predicate $\tyface{is\_true}\,v$ 
holds if $v$ is a pointer or a nonzero integer; 
$ \tyface{is\_false}$ holds only on 0.
A store statement $\Sstore{\mathit{ch}}{e_1}{e_2}$ requires the corresponding 
store permission $\infootprint{v_1}{\mathit{ch}}{store}{\phi_\sigma} $.

Given a control stack $\Kseq {\Sblock s}\ctl$, the small-step execution of the block 
statement $\Sblock s$ enters that block: $s$ becomes the next statement to execute and the
control stack becomes ${\Kseq{s}{\Kblock\ctl}}$.

Exit statements are only allowed from blocks that have been previously entered.
For that reason, in the two rules for exit statements, the control stack ends with
$(\Kblock \ctl)$ control.
A statement $(\Sexit n)$ terminates the $(n+1)^{th}$ enclosing block statements.
In such a block, the stack of control sequences ${s_1}\cdots{s_j}$ following the
exit statement is not executed. 
Let us note that this stack may be empty if the exit statement is the last statement
of the most enclosing block.
The small-step execution of a statement $(\Sexit n)$
exits from only one block (the most enclosing one).
Thus, the execution of an $(\Sexit 0)$ statement updates the control stack
$(\Kseq{\Sexit{0}}{\Kseq{s_1}{\cdots\Kseq{s_j}{\Kblock{\ctl}}}})$ into $\ctl$.
The execution of an $(\Sexit {\,n+1})$ statement updates the control stack
$(\Kseq{\Sexit{(n+1)}}{\Kseq{s_1}{\cdots\Kseq{s_j}{\Kblock{\ctl}}}})$ into $\Kseq{\Sexit n}\ctl$.

%
%

\begin{figure}[t]
\begin{mathpar}
\step{\Cont{\sigma}{\,\Kseq{(\Sseq{s_1}{s_2})}{\ctl}}}{\Cont{\sigma}{
\Kseq{s_1}{\Kseq{s_2}{\ctl}}}}
\and
\infrencee{\evalexp{\sigma}{e}{v} \quad
\rho' = \rho_\sigma[x:=v]}
{\step{\Cont{\sigma}{\,\Kseq {(\Sassign{x}{e})} \ctl}}
{\Cont{\sigma[:=\rho']}{\ctl}}}
\and
\infrencee{
\evalexp{\sigma}{e_1}{v_1} \and \evalexp{\sigma}{e_2}{v_2}  \and
\infootprint{v_1}{\mathit{ch}}{store}{\phi_\sigma} \and
\storev{\mathit{ch}}{m_\sigma}{v_1}{v_2}{m'}
}{
\step{\Cont{\sigma}{\,\Kseq {(\Sstore{\mathit{ch}}{e_1}{e_2})} \ctl}}
{\Cont{\sigma[:=m']}{\ctl}}}
\and
\infrencee{
\evalexp{\sigma}{e}{v}\and\qquad \tyface{is\_true}\,v}
{\step{\Cont{\sigma}{\,\Kseq{(\Sif{e}{s_1}{s_2})}{\ctl}}}{\Cont{\sigma}{\Kseq{s_1}{\ctl}}}}\and
\infrencee{
\evalexp{\sigma}{e}{v}\and\qquad \tyface{is\_false}\,v}
{\step{\Cont{\sigma}{\,\Kseq{(\Sif{e}{s_1}{s_2})}{\ctl}}}{\Cont{\sigma}{\Kseq{s_2}{\ctl}}}}\and
\step{\Cont{\sigma}{\Kseq{\Sskip}{\ctl}}}
{\Cont{\sigma}{\ctl}}
\and
\step{\Cont{\sigma}{\,\Kseq{(\Sloop{s})}{\ctl}}}
{\Cont{\sigma}{\Kseq{s}{\Kseq{\Sloop{s}}{\ctl}}}}
\and
\step{\Cont{\sigma}{\Kseq{(\Sblock{s})}{\ctl}}}
{\Cont{\sigma}{\Kseq{s}{\Kblock\ctl}}}
\and
\infrencee{
j\ge 1
}
{\step{\Cont{\sigma}{\Kseq{\Sexit{0}}{\Kseq{s_1}{\cdots\Kseq{s_j}{\Kblock{\ctl}}}}}}
{\Cont{\sigma}{\ctl}}}
\and
\infrencee{
j\ge 1
}
{\step{\Cont{\sigma}{\Kseq{\Sexit{(n+1)}}{\Kseq{s_1}{\cdots\Kseq{s_j}{\Kblock{\ctl}}}}}}
{\Cont{\sigma}{\Kseq{\Sexit{n}}{\ctl}}}}
\end{mathpar}
\caption{Sequential small-step relation.  We omit here call and return, which are in the full technical report \cite{appel07:tr}.}
\label{fig-step}
\end{figure}

\begin{lemma}
If $\evalexp{\sigma}{e}{v}$
then 
$\step{\Cont\sigma{\Kseq{(\Sassign{x}{e})}\ctl}}{k'}$
iff
$\step{\Cont\sigma{\Kseq{(\Sassign{x}{\Eval\,v})}\ctl)}}{k'}$
(and similarly for other statements containing expressions).
\vspace{-7pt}
\begin{proof}
Trivial: expressions have no side effects.
A convenient property nonetheless, and not true of 
Leroy's original \cminor{}.
$\blacksquare$
\end{proof}
\end{lemma}

\begin{definition}
A continuation $k=(\sigma,\ctl)$ 
is \tyface{stuck} if
$\ctl \not=\Kstop$ and
there does not exist $k'$ such that 
$ \step{k}{k'}$.
\end{definition}

\begin{definition}\label{def:safe}
A continuation $k$ is \tyface{safe} (written as $\fmap\vdash\tyface{safe}(k)$\label{lab:safe}) if
it cannot reach a stuck continuation
in the sequential small-step relation $\longmapsto^{*}$.
\end{definition}

\section{Separation Logic}
\label{seplog}

Hoare Logic uses triples $\{P\}\,s\,\{Q\}$ where
$P$ is a precondition, $s$ is a statement of the programming
language, and $Q$ is a postcondition. The assertions $P$ and $Q$ are
predicates on the program state. 
The reasoning on memory is inherently global.
Separation Logic is an extension of Hoare Logic for programs that manipulate pointers.
In Separation Logic, reasoning is local~\cite{ohearn01};
assertions such as $P$ and $Q$ describe properties of part of the memory, 
and $\{P\}\,s\,\{Q\}$
describes changes to part of the memory.
We prove the soundness of the Separation Logic
via a shallow embedding, that is, we give
each assertion a semantic meaning in Coq.
We have
$P,Q: \tyface{assert}$~~~where~~ $\tyface{assert}=\tyface{prog}\rightarrow\tyface{state}\rightarrow\tyface{Prop}$.
So $P\fmap\sigma$ is a proposition of logic and we say that $\sigma$ satisfies $P$.

\paragraph{Assertion Operators.}

\begin{figure}
\vspace{-30pt}
\begin{align*}
\emp =_\mathrm{def}~ & \lambda\fmap\sigma.~\phi_{\sigma}=\emptyset \\
P*Q =_\mathrm{def}~ & \lambda\fmap\sigma.~
\exists \phi_1. \exists \phi_2.~~
 \phi_\sigma=\phi_1\oplus\phi_2
~\wedge~  P(\sigma[:=\phi_1])~\wedge~Q(\sigma[:=\phi_2]) \\
P\vee Q =_\mathrm{def}~ & \lambda\fmap\sigma.~P\sigma\,\vee\,Q\sigma \\
P\wedge Q =_\mathrm{def}~ & \lambda\fmap\sigma.~P\sigma\,\wedge\,Q\sigma \\
P\Rightarrow Q =_\mathrm{def}~ & \lambda\fmap\sigma.~P\sigma\,\Rightarrow\,Q\sigma \\
\neg P =_\mathrm{def}~ & \lambda\fmap\sigma.~\neg (P\sigma)\\
\exists z.P =_\mathrm{def}~ & \lambda\fmap\sigma.~\exists z.\,P\sigma \\
\assertprop{A}=_\mathrm{def}~ &\lambda\fmap\sigma.~A \qquad \mbox{\small where $\sigma$
does not appear free in $A$} \\
\TT =_\mathrm{def}~ & \lambda\fmap\sigma.\mathbf{True} \qquad \qquad \qquad
\FF =_\mathrm{def}~ \assertprop{\mathrm{\mathbf{False}}} \\
e \Downarrow v ~=_\mathrm{def}~&  \emp\,\wedge\,
\assertprop{\tyface{pure}(e)}\,\wedge\, 
 \lambda\fmap\sigma.
\,(\evalexp{\sigma}{e}{v}) \\
\assertexpr{e}=_\mathrm{def} ~& \exists v.\,e\Downarrow v\, \wedge\,
\assertprop{\tyface{is\_true}\,v} \\
\assertdef{e} =_\mathrm{def}~&
\assertexpr{\ceq{\mathrm{int}}{e}{e}}\,\vee\,
\assertexpr{\ceq{\mathrm{float}}{e}{e}} \\
\assertmapsto{e_1}{\mathit{ch}}{e_2}  =_\mathrm{def}~&
\exists v_1. \exists v_2. (e_1 \Downarrow v_1) \wedge
(e_2 \Downarrow v_2) \wedge
(\lambda \sigma, \loadv{m_\sigma}{v_1}{\mathit{ch}}{v_2} \wedge \infootprint{v_1}{\mathit{ch}}{store}{\phi_{\sigma}}) \wedge
\assertdef{v_2}
\end{align*}
\vspace{-20pt}
\caption{Main operators of Separation Logic}
\label{seplogfig}
\vspace{-10pt}
\end{figure}

In Fig.~\ref{seplogfig}, we define the usual operators of Separation Logic:
the empty assertion $\emp$, separating conjunction $*$, 
disjunction $\vee$, conjunction $\wedge$, implication $\Rightarrow$,
negation $\neg$, and quantifier $\exists$.
A state $\sigma$ satisfies $P*Q$ if its footprint $\phi_\sigma$ can be split into $\phi_1$ and $\phi_2$
such that $\sigma[:=\phi_1]$ satisfies $P$ and $\sigma[:=\phi_2]$ satisfies $Q$.
We also define some novel operators such as
expression evaluation $e\Downarrow v$
and base-logic propositions  $\assertprop{A}$.

O'Hearn and Reynolds specify Separation Logic for a little
language in which expressions evaluate independently of the heap~\cite{ohearn01}.
That is, their expressions access only the program variables
and do not even have \emph{read} side effects on the memory.
Memory reads
are done by a command of the language, not within
expressions.  In \cminor{} we relax this restriction; expressions
can read the heap.  But we say that an expression is \emph{pure} if
it contains no \tyface{Eload} operators---so that it cannot read the heap.

In Hoare Logic one can use expressions of the programming language as
assertions---there is an implicit coercion.
We write the assertion $e\Downarrow v$ 
to mean that
expression $e$ is pure and evaluates to value $v$ in the operational semantics.
This is an expression of Separation Logic, in contrast to 
$\evalexp{\sigma}{e}{v}$ which is a judgment in the underlying
logic.
In a previous experiment, our Separation Logic permitted impure expressions
in $e\Downarrow v$. But, this complicated the proofs unnecessarily. Having 
$ \emp\,\wedge\,\assertprop{\tyface{pure}(e)}$ in the definition of $e\Downarrow v$
leads to an easier-to-use Separation Logic.

Hoare Logic traditionally allows expressions $e$ of the programming
language to be used as expressions of the program logic.
We will define explicitly
$\assertexpr{e}$ to mean that $e$ evaluates to a true value
(\textit{i.e.} a nonzero integer or non-null pointer).
Following Hoare's example, we will usually omit the $\assertexpr{\,}$
braces in our Separation Logic notation.

\cminor's integer equality operator,
which we will write as $\ieq{e_1}{e_2}$,
applies to integers or pointers, but in several cases it is
``stuck'' (expression evaluation gives no result):
when comparing a nonzero integer to a pointer;
when comparing
\Vundef{} or $\Vfloat{(x)}$ to anything.  Thus we can write
the assertion $\assertexpr{\ieq e e}$ (or just write $\ieq e e$) to test 
that $e$ is a defined integer or pointer in the current state,
and there is a similar operator $\ceq{\mathrm{float}}{e_1}{e_2}$.

Finally, we have the usual Separation Logic singleton ``maps-to'',
but annotated with a chunk-type $\mathit{ch}$.  That is,
$\assertmapsto{e_1}{\mathit{ch}}{e_2}$
means that $e_1$ evaluates to $v_1$, $e_2$ evaluates to $v_2$, and
at address $v_1$ in memory there is a defined value
$v_2$ of the given chunk-type.
Let us note that in this definition, $\assertdef{v_1}$ is implied by the third conjunct.
$\assertdef{v_2}$ is a design decision. We could leave it out and have a slightly different
Separation Logic.


\paragraph{The Hoare Sextuple.}
\cminor{} has commands to
call functions, to \tyface{exit} (from a block), and to
\tyface{return} (from a function).  Thus, we extend the 
Hoare triple $\{P\}\,s\,\{Q\}$ with three extra contexts to become
$\,\semax{\Gamma}{R}{B}{P}{s}{Q}$
where:\vspace{-7 pt}
\begin{itemize}
\item[$\Gamma:$]$\tyface{assert}$ describes
context-insensitive properties of the global environment;
\item[$R:$]$\tyface{list}\,\tyface{val}\!\rightarrow\!\tyface{assert}$
is the \emph{return environment}, 
giving the current function's postcondition as a predicate
on the list of returned values; and 
\item[$B:$]$ \tyface{nat}\!\rightarrow\!\tyface{assert}$ is the
\emph{block environment} giving the exit conditions of each \Sblock statement
in which the statement $s$ is nested.
\end{itemize}

Most of the rules of sequential Separation Logic are given
in Fig.~\ref{seq-sep-fig}. In this paper, we omit the rules for return and call, 
which are detailed in the full technical report. Let us note that the $\Gamma$ context
is used to update global function names, none of which is illustrated in this paper.

\begin{figure}[th]
\vspace{-21pt}
\begin{mathpar}
\inference{
P\Rightarrow P' \and \semax{\Gamma}{R}{B}{P'}{s}{Q'}
\and Q'\Rightarrow Q
}{
\semax{\Gamma}{R}{B}{P}{s}{Q}
}
\and
\semax{\Gamma}{R}{B}{P}{\Sskip}{P}
\and
\inference{
\semax{\Gamma}{R}{B}{P}{s_1}{P'}
\and \semax{\Gamma}{R}{B}{P'}{s_2}{Q}
}{
\semax{\Gamma}{R}{B}{P}{\Sseq{s_1}{s_2}}{Q}
}
\and
\inference{
\rho' = \rho_\sigma[x:=v] \and
P=(\exists v. \,
e\Downarrow v ~\wedge~
\lambda\st.~Q\,\sigma[:={\rho'}])
}{
\semax{\Gamma}{R}{B}{P}{\Sassign{x}{e}}{Q}
}
\and
\inference{
\pure(e)\and \pure(e_2)\and
P=(\assertmapsto{e}{\mathit{ch}}{e_2}~ \wedge ~
\assertdef{e_1})
}{
\semax{\Gamma}{R}{B}{P}{\Sstore{\mathit{ch}}{e}{e_1}}{
\assertmapsto{e}{\mathit{ch}}{e_1}}
}
\and
\inference{
  \pure (e)\and
  \semax{\Gamma}{R}{B}{P\wedge e}{s_1}{Q}\and
\semax{\Gamma}{R}{B}{P\wedge \neg e}{s_2}{Q}
}{
\semax{\Gamma}{R}{B}{P}{\Sif{e}{s_1}{s_2}}{Q}
}
\and
\inference{
\semax{\Gamma}{R}{B}{I}{s}{I}
}{
\semax{\Gamma}{R}{B}{I}{\Sloop{s}}{\FF}
}
\and
\inference{
\semax{\Gamma}{R}{\Bcons{Q}{B}}{P}{s}{\FF}  
}{
\semax{\Gamma}{R}{B}{P}{\Sblock{s}}{Q}}
\and
\semax{\Gamma}{R}{B}{B(n)}{\Sexit{n}}{\FF}
\and
\inference{
\semax{\Gamma}{R}{B}{P}{s}{Q}
\and
\mathrm{modified\,vars}(s)~\cap~\mathrm{free\,vars}(A)~=~\emptyset
}{
\semax{\Gamma}{(\lambda \mathit{vl}. A* R(\mathit{vl}))}{(\lambda n.A*B(n))}{A * P}{s}{A*Q}
}
\end{mathpar}
\vspace{-10pt}
\caption{Axiomatic Semantics of Separation Logic (without call and return)}
\label{seq-sep-fig}
\vspace{-10pt}
\end{figure}

The rule for ${\Sstore{\mathit{ch}}{e}{e_1}}$ requires the same store permission 
than the
small-step rule, but in Fig.~\ref{seq-sep-fig}, the permission is hidden in the definition of
$\assertmapsto{e}{\mathit{ch}}{e_2}$.
The rules for ${\Sstore{\mathit{ch}}{e}{e_1}}$ and $\Sif{e}{s_1}{s_2}$ require that
$e$ be a pure expression.  To reason about an such statements
where $e$ is impure, one reasons by program transformation
using the following rules. It is not necessary to rewrite the actual source
program, it is only the local reasoning that is by program 
transformation.
\vspace{-2pt}
\begin{mathpar}
\inference{
x,y~\mbox{not free in}~e, e_1,Q \qquad \qquad
\semax{\Gamma}{R}{B}{P}{\,\Sseq{\Sassign{x}{e}}{\,\Sseq{\Sassign{y}{e_1}}{\,\Sstore{\mathit{ch}}{x}{y}}}\,}{Q}}
{
\semax{\Gamma}{R}{B}{P}{\Sstore{\mathit{ch}}{e}{e_1}}{Q}
}
\and
\inference{
x~\mbox{not free in}~s_1,s_2,Q \qquad \qquad
\semax{\Gamma}{R}{B}{P}{\,\Sseq{\Sassign{x}{e}}{\Sif{x}{s_1}{s_2}}\,}{Q}}
{\semax{\Gamma}{R}{B}{P}{\,\Sif{e}{s_1}{s_2}\,}{Q}}
\end{mathpar}

The statement $\Sexit{i}$ exits from the $(i+1)^{th}$ enclosing
\tyface{block}.
A block environment $B$ is a sequence of assertions
$B_0,B_1,\ldots,B_{k-1}$ such that $(\Sexit{i})$ is safe
as long as the precondition $B_i$ is satisfied.  We
write $\Bnil$ for the empty block environment
and $B'=\Bcons{Q}{B}$ for the environment such that
$B'_0=Q$ and $B'_{i+1}=B_i$.

Given a block environment $B$, a precondition $P$ and a postcondition $Q$,
the axiomatic semantics of a $(\Sblock s)$ statement consists in executing
some statements of
$s$ given the same precondition $P$ and the block environment $\Bcons{Q}{B}$
(\textit{i.e.} each existing block nesting is incremented).
The last statement of $s$ to be executed is an exit statement that yields the
$\FF$ postcondition.
An $(\Sexit n)$ statement is only allowed from a corresponding enclosing block,
\textit{i.e.} the precondition $B(n)$ must exist in the block environment $B$
and it is the precondition of the $(\Sexit n)$ statement.



\paragraph{Frame Rules.}
The most important feature of Separation Logic is the frame rule, usually written
\vspace{-6pt}
\[
\inference{
\{P\}{\,s\,}\{Q\}
}{
\{A\,*P\}{\,s\,}\{A\,*Q\}
}
\] 
The appropriate generalization of this rule to our language with control flow
is the last rule of Fig.~\ref{seq-sep-fig}. 
We can derive from it a \emph{special frame rule}
for simple statements $s$ that
do not exit or return:
\[
\inference{
\forall R,B.(\semax{\Gamma}{R}{B}{P}{\,s\,}{Q})
\qquad
\mathrm{modified\,vars}(s)~\cap~\mathrm{free\,vars}(A)~=~\emptyset
}{
\semax{\Gamma}{R}{B}{A*P}{\,s\,}{A*Q}
}
\]

\paragraph{Free Variables.}
We use a semantic notion of free variables: $x$ is not free in
assertion $A$ if, in any two states where only the binding of $x$
differs, $A$ gives the same result.  However, we found it necessary to
use a syntactic (inductive) definition of the variables modified by a
command.  One would think that command $c$ ``modifies'' $x$ if there
is some state such that by the time $c$ terminates or exits, $x$ has a
different value.  However, this definition means that the modified
variables of \texttt{if~false~then~$B$~else~$C$} are \emph{not} a
superset of the modified variables of $C$; this lack of an inversion
principle led to difficulty in proofs.

\paragraph{Auxiliary Variables.}  It is typical in Hoare Logic to use auxiliary variables to relate the pre- and postconditions, e.g., the variable $a$ in
$\{x=a\}\,x:=x+1\,\{x=a+1\}$.  In our shallow embedding of Hoare Logic
in Coq, the variable $a$ is a Coq variable, not a \cminor{} variable;
formally, the user would prove in Coq the proposition, $\forall
a,(\semax{\Gamma}{R}{B}{P}{s}{Q})$ where $a$ may appear free in any of
$\Gamma,R,B,P,s,Q$.  The existential assertion $\exists z.Q$ is useful
in conjunction with this technique.

Assertions about functions require special handling of these 
quantified auxiliary variables.  The assertion that some value $f$ is a 
function with precondition $P$ and postcondition $Q$ is written
$f: \underline{\forall} x_1\underline{\forall} x_2\ldots\underline{\forall} x_n,\,\{P\}\{Q\}$
where $P$ and $Q$ are functions from value-list to assertion, each
$\underline{\forall}$ is an operator of our separation logic that
binds a Coq variable $x_i$ using higher-order abstract syntax.

\paragraph{Application.}
In the full technical report \cite{appel07:tr}, 
we show how the Separation Logic (\textit{i.e.} the rules of Fig.~\ref{seq-sep-fig}) 
can be used to prove partial correctness properties of programs,
with the classical in-place list-reversal example.
Such proofs rely on a set of tactics, that we have written in the tactic definition language of Coq, 
to serve as a proof assistant for \cminor{} Separation Logic proofs~\cite{appel06:septacs}.

\section{Soundness of Separation Logic}\label{sec:soundness}

Soundness means not only that there is a model for the
logic, but that the model is \emph{the} operational semantics 
for which the compiler guarantees correctness!
In principle we could prove soundness by syntactic induction over the 
Hoare Logic rules, but instead we will give a semantic definition
of the Hoare sextuple $\semax{\Gamma}{R}{B}{P}{\,s\,}{Q}$,
and then prove each of the Hoare rules as a derived lemma from 
this definition.

A simple example of semantic specification is that the
Hoare Logic $P\Rightarrow Q$ is defined, using the
underlying logical implication, as 
$\forall \fmap \sigma.~P\,\fmap\,\sigma~\Rightarrow~ Q\,\fmap\,\sigma$.
From this, one could prove soundness of the Hoare Logic rule
on the left (where the $\Rightarrow$ is a symbol of Hoare Logic)
by expanding the definitions into the lemma on the right
(where the $\Rightarrow$ is in the underlying logic),
which is clearly provable in higher-order logic:
\vspace{-2pt}
\[\inference{P\Rightarrow Q\qquad Q \Rightarrow R}{P \Rightarrow R}
\qquad
\inference{\forall \fmap \sigma.(P\fmap\sigma \Rightarrow Q\fmap\sigma )
\quad \forall \fmap\sigma.(Q\fmap\sigma  \Rightarrow R\fmap\sigma )}
{\forall \fmap\sigma.(P\fmap\sigma  \Rightarrow R\fmap\sigma )}\]

\begin{definition}
(a) Two states $\sigma$ and $\sigma'$ are \tyface{equivalent} 
(written as $\sigma\cong\sigma'$) if they have the same stack pointer, 
extensionally equivalent environments, identical footprints, and if
the footprint-visible portions of their memories are the same.
(b) An \tyface{assertion} is a predicate on states that is 
extensional over equivalent environments (in Coq it is a dependent
product of a predicate and a proof of extensionality).
\end{definition}

\begin{definition}
For any control $\ctl$, we define the 
assertion $\mathbf{safe}~\ctl$ to mean that the combination of $\ctl$ with
the current state is safe:
\[\mathbf{safe}~\ctl~=_\mathrm{def}~\lambda \fmap \sigma.\,\forall \sigma'.~
(\sigma\cong\sigma'\,\Rightarrow~\fmap\vdash \safe(\sigma',\kappa))\]
\end{definition}

\begin{definition}
Let $A$ be a \emph{frame}, that is, a closed assertion
(\textit{i.e.} one with no free \cminor{} variables).
An assertion $P$ \tyface{guards} 
a control ${\ctl}$ in the frame $A$
(written as $\guard{A}{P}{\ctl}$)
means that whenever $A*P$ holds, it is safe to execute $\ctl$.
That is,
\[
\guard{A}{P}{\ctl}~=_\mathrm{def}~ A*P\, \Rightarrow\, \mathbf{safe}~\ctl.
\]
We extend this notion to say that a return-assertion $R$
(a function from value-list to assertion) guards a return,
and a block-exit assertion $B$ (a function from block-nesting level to
assertions) guards an exit:
\begin{mathpar}
\rguard{A}{R}{\ctl}~\defeq ~\forall \mathit{vl}. \guard{A}{R(\mathit{vl})}{\returncont{\mathit{vl}}{\ctl}}
 \qquad  \qquad 
\bguard{A}{B}{\ctl}~\defeq ~\forall n. \guard{A}{B(n)}{\exitcont{\mathit{n}}{\ctl}} 
\end{mathpar}
\end{definition}

\begin{lemma}
\label{lemma-seq1}
If $\guard{A}{P}{\Kseq {s_1} \Kseq {s_2} \ctl}$ then
$\guard{A}{P}{\Kseq {(\Sseq {s_1} {s_2})} \ctl}$.
\end{lemma}

\begin{lemma}
\label{lemma-seq2}
If $\rguard{A}{R}{\ctl}$ then
$\forall s,\rguard{A}{R}{\Kseq{s}\ctl}$.~~~~~~~~If $\bguard{A}{B}{\ctl}$ then
$\forall s,\bguard{A}{B}{\Kseq{s}\ctl}$.
\end{lemma}

\begin{definition}[Frame]
A \tyface{frame} is constructed from the global environment $\Gamma$,
an arbitrary frame assertion $A$, and a statement $s$, by
the conjunction of $\Gamma$ with the assertion $A$ closed over any 
variable modified by $s$:
\[\tyface{frame}(\Gamma,A,s)~\defeq ~
\Gamma * \tyface{closemod}(s,A)\]
\end{definition}

\begin{definition}[Hoare sextuples]
The Hoare sextuples are defined in ``continuation style,'' in
terms of implications between continuations, as follows:
\vspace{-5pt}
\begin{mathpar}
\boxed{
\begin{array}{l}
\semax{\Gamma}{R}{B}{P}{\,s\,}{Q} ~\defeq ~ ~\forall A,\ctl.~~
\\
~~~\rguard{\tyface{frame}(\Gamma,A,s)}{R}{\ctl}\,\wedge
\,\bguard{\tyface{frame}(\Gamma,A,s)}{B}{\ctl}\,\wedge\,
\guard{\tyface{frame}(\Gamma,A,s)}{Q}{\ctl}~ 
\Rightarrow ~\guard{\tyface{frame}(\Gamma,A,s)}{P}\Kseq{s}{\ctl}~
\end{array}}
\end{mathpar}
\end{definition}
From this definition we prove the rules of Fig.~\ref{seq-sep-fig}
as derived lemmas.

It should be clear from the definition---after one gets over the 
backward nature of the continuation transform---that
the Hoare judgment specifies
partial correctness, not total correctness.
For example, if the statement $s$
infinitely loops, then the continuation $(\st,\Kseq{s}{\ctl})$ is automatically
safe, and therefore $\guard{A}{P}{\Kseq{s}{\ctl}}$ always holds.  Therefore
the Hoare tuple $\semax{\Gamma}{R}{B}{P}{s}{Q}$ will hold for that $s$,
regardless of $\Gamma,R,B,P,Q$.

\paragraph{Sequence.}
The soundness of the sequence statement is the proof that 
if the hypotheses $\tyface{H_1}:\semax{\Gamma}{R}{B}{P}{\,s_1\,}{P'}$ and
$\tyface{H_2}:\semax{\Gamma}{R}{B}{P'}{\,s_2\,}{Q}$ hold, then we have to prove
$\tyface{Goal}:\semax{\Gamma}{R}{B}{P}{\,\Sseq{s_1}{s_2}\,}{Q}$ (see Fig.~\ref{seq-sep-fig}).
If we unfold the definition of the Hoare sextuples, 
$\tyface{H_1}$, $\tyface{H_2}$ and $\tyface{Goal}$ become:

\newcommand{\negsp}{\hspace{-8pt}}
\begin{mathpar}
\inference[($\forall A,\ctl_i$)]
{\negsp{\rguard{\tyface{frame}(\Gamma,A,s_i)}{R}{\ctl_i}} \and \bguard{\tyface{frame}(\Gamma,A,s_i)}{B}{\ctl_i} \negsp \and
\negsp \guard{\tyface{frame}(\Gamma,A,s_i)}{P'}{\ctl_i} \negsp}
{\guard{\tyface{frame}(\Gamma,A,s_i)}{P}{\Kseq{s_i}{\ctl_i}}}[$\tyface{H}_i, i=1,2$]
\and 
\inference[($\forall A,\ctl$)]
{{\rguard{\tyface{frame}(\Gamma,A,(s_1;s_2))}{R}{\ctl}} \and \bguard{\tyface{frame}(\Gamma,A,(s_1;s_2))}{B}{\ctl} \and
\guard{\tyface{frame}(\Gamma,A,(s_1;s_2))}{Q}{\ctl}}
{\guard{\tyface{frame}(\Gamma,A,(s_1;s_2))}{P}{\Kseq{(\Sseq{s_1}{s_2})}{\ctl}}}[$\tyface{Goal}$]
\end{mathpar}

We prove $\guard{\tyface{frame}(\Gamma,A,(s_1;s_2))}{P}{\Kseq{(\Sseq{s_1}{s_2})}{k}}$ using
Lemma~\ref{lemma-seq1}:\footnote{
We will elide the frames from proof sketches by writing 
$\guard{}{}{}$ without a subscript; this particular
proof relies on a lemma that 
\(\tyface{closemod}(s_1,\tyface{closemod}((s_1;s_2),A))=\tyface{closemod}((s_1;s_2),A)\).}
\begin{mathpar}
\inference{
\hspace{-8pt}
\inference
{\rguard{}{R}{k}
}
{\rguard{}{R}{\Kseq{s_2}{k}}}[\textsf{\hspace{-5pt}Lm.~\ref{lemma-seq2}}]
\and
\hspace{-8pt}
\inference
{\bguard{}{B}{k}
}
{\bguard{}{B}{\Kseq{s_2}{k}} }[\textsf{\hspace{-5pt}Lm.~\ref{lemma-seq2}}]
\and
\hspace{-8pt}
\inference
{\rguard{}{R}{k} ~~~
\bguard{}{B}{k}~~~ \guard{}{Q}{k}
}
{\guard{}{P'}{\Kseq{s_2}{k}}}[\hspace{-5pt}$\tyface{H_2}$]
\hspace{-8pt}
}
{\inference
{\guard{}{P}{\Kseq{s_1}{\Kseq{s_2}{k}}}
}{\guard{}{P}{\Kseq{(\Sseq{s_1}{s_2})}{k}} }[\textsf{Lm.~\ref{lemma-seq1}}]}[$\tyface{H_1}$]
\end{mathpar}

\paragraph{Loop Rule.}
The loop rule turns out to be one of the most difficult ones to prove.
A loop continues executing until the loop-body performs an \tyface{exit}
or \tyface{return}.  If $\Sloop{s}$ executes $n$ steps, then
there will be 0 or more complete iterations of $n_1, n_2,\ldots$ 
steps, followed by $j$ steps into the last iteration.
Then either there is an exit (or return) from the loop, or the loop
will keep going.  But if the \tyface{exit} is from an inner-nested
\tyface{block}, then it does not terminate the loop (or even this
iteration).  Thus we need a formal notion of when a statement
exits.

Consider the statement $s=\Sif{b}{\Sexit{2}}{(\Sseq{\Sskip{}}{\Sassign{x}{y}})}$, executing in state $\st$.
Let us execute $n$ steps into $s$, that is, 
$\stepn{(\st,\Kseq{s}{\ctl})}{(\st',\ctl')}{n}$.
If $n$ is small, then the behavior should not depend on $\ctl$;
only when we ``emerge'' from $s$ is $\ctl$ important.
In this example,
if $\rho_\st b$ is a true value, then as long as $n\le 1$ the
statement $s$ can \emph{absorb} $n$ steps independent of $\ctl$;
if $\rho_\st b$ is a false value, then $s$ can absorb up to 3 steps.
To reason about absorption, we define the
concatenation $\cat{\ctl_1}{\ctl_2}$ 
of a control prefix $\ctl_1$ and a control $\ctl_2$ as follows:
\vspace{-7pt}
\[
\begin{array}{rclrcl}
\cat{\Kstop}{\ctl} &=_\mathrm{def}& \ctl &
\cat{(\Kblock{\ctl'})}{\ctl} &=_\mathrm{def}& \Kblock{(\cat{\ctl'}{\ctl})}\\
\cat{(\Kseq{s}{\ctl'})}{\ctl} &=_\mathrm{def}& \Kseq{s}{(\cat{\ctl'}{\ctl})}\,\qquad\qquad&
\cat{(\Kcall{\mathit{xl}}{f}{\mathit{sp}}{\rho}{\ctl'})}{\ctl} &=_\mathrm{def}& 
\Kcall{\mathit{xl}}{f}{\mathit{sp}}{\rho}{(\cat{\ctl'}{\ctl})}\\
\end{array}\]
$\Kstop$ is the empty prefix; 
~~$\cat{\Kstop}{\ctl}$ does not mean ``stop,'' it means $\ctl$.

\begin{definition}[absorption]
A statement $s$ in state $\st$ \tyface{absorbs} $n$ steps (written as  $\absorb{n}{s}{\st}$) iff 
$\forall j \leq n.\, \exists \ctl_{\mathrm{prefix}}.\exists \st'.~ \forall \ctl.~
\stepn{\Cont{\st}{\Kseq{s}{\ctl}}}{\Cont{\st'}{\cat{\ctl_{\mathrm{prefix}}}{\ctl}}}{j}$.
\end{definition}

\begin{example}
An \tyface{exit} statement by itself absorbs no steps (it immediately
uses its control-tail), but
$\Sblock{(\Sexit 0)}$ can absorb the 2 following steps:\newline
$\fmap ~\vdash ~ (\st,\Kseq{\Sblock{(\Sexit 0)}}{\ctl})\longmapsto
(\st,\Kseq{\Sexit 0}{\Kblock\ctl})\longmapsto
(\st,\ctl)$
\end{example}

\begin{lemma}
\begin{enumerate}
\item $\absorb{0}{s}{\st}$.
\item $\absorb{n+1}{s}{\st}~ \Rightarrow~ \absorb{n}{s}{\st}$.
\item If $\neg \absorb {n}{s}{\st}$, then 
$\exists i<n. \absorb{i}{s}{\st} \wedge \neg \absorb{i+1}{s}{\st}$.
We say that $s$ absorbs at most $i$ steps in state $\st$.
\end{enumerate}
\end{lemma}


\begin{definition}
We write $(s;)^n s'$ to mean
$\underbrace{s; s ; \ldots ; s ; {}}_{n} s'$.
\end{definition}
\vspace{-8pt}
\vspace{-7pt}
\begin{lemma}
\label{semax_loop2}
\(
\inferrule{\semax{\Gamma}{R}{B}{I}{s}{I}}
{\semax{\Gamma}{R}{B}{I}{(s;)^n \Sloop{\Sskip}}{\FF}}
\)
\begin{proof}
For $n=0$, the infinite-loop $(\Sloop{\Sskip})$ satisfies any precondition for
partial correctness.
For $n+1$, 
assume $\ctl$, 
$\rguard{}{R}{\ctl}$, 
$\bguard{}{B}{\ctl}$;
by the induction hypothesis (with $\rguard{}{R}{\ctl}$ and $\bguard{}{B}{\ctl}$)
we know $\guard{}{I}{(s;)^n\Sloop{\Sskip}\cdot\ctl}$.
We have 
$\rguard{}{R}{(s;)^n\Sloop{\Sskip}\cdot \ctl}$ and
$\bguard{}{B}{(s;)^n\Sloop{\Sskip}\cdot \ctl}$
by Lemma~\ref{lemma-seq2}.
We use the hypothesis $\semax{\Gamma}{R}{B}{I}{s}{I}$ to augment the result to
$\guard{}{I}{(s;(s;)^n\Sloop{\Sskip})\cdot\ctl}$. 
$\blacksquare$
\end{proof}
\end{lemma}

\begin{theorem}
\label{semax_loop1}
\(
\inferrule{\semax{\Gamma}{R}{B}{I}{s}{I}}
{\semax{\Gamma}{R}{B}{I}{\Sloop{s}}{\FF}}
\)
\begin{proof}
Assume $\ctl$, $\rguard{}{R}{\ctl}$, 
$\bguard{}{B}{\ctl}$.
To prove $\guard{}{I}{\Kseq{\Sloop{s}}{\ctl}}$,
assume $\st$ and $I\st$ and prove
$\safe{\Cont{\st}{\Kseq{\Sloop s}{\ctl}}}$.
We must prove that for any $n$, after $n$ steps we are not stuck.
We unfold the loop $n$ times, that is,
we use Lemma~\ref{semax_loop2} to show
$\safe{\Cont{\st}{(s ;)^n \Sloop{\Sskip} \cdot \ctl}}$.
We can show that if this is safe for $n$ steps,
so is $\Kseq{\Sloop s}{\ctl}$ by the principle of 
absorption.  Either $s$ absorbs $n$ steps, in which case we are done;
or $s$ absorbs at most $j<n$ steps, leading to a state $\st'$
and a control (respectively)
$\ctl_\textrm{prefix} \circ (s ;)^{n-1} \Sloop{\Sskip} \cdot \ctl$
or 
$\ctl_\textrm{prefix} \circ \Sloop{s} \cdot \ctl$.
Now, because $s$ cannot absorb $j+1$ steps,
we know that either $\ctl_\textrm{prefix}$ is empty
(because $s$ has terminated normally) or
$\ctl_\textrm{prefix}$ starts with a \tyface{return} or \tyface{exit},
in which case we escape (resp. past the $\Sloop{\Sskip}$
or the $\Sloop{s}$) into $\ctl$.
If $\ctl_\textrm{prefix}$ is empty then we apply strong induction
on the case $n-j$ steps; if we escape, then 
$\Cont{\st'}{\ctl}$ is safe iff 
$\Cont{\st}{\Kseq{\Sloop s}{\ctl}}$ is safe.
(For example, if $j=0$, then it must be that $s=\Sreturn$ or $s=\Sexit$, 
so in one step we reach $\ctl_\textrm{prefix} \circ (\Sloop{s} \cdot \ctl)$
with $\ctl_\textrm{prefix} = \Sreturn$ or $\ctl_\textrm{prefix} = \Sexit$.)
$\blacksquare$
\end{proof}
\end{theorem}

\section{Sequential Reasoning about Sequential Features}

Concurrent \cminor, like most concurrent programming languages
used in practice, is a sequential programming language
with a few concurrent features (locks and threads)
added on.  We would like to be able to reason about the sequential
features using purely sequential reasoning.   If we have to
reason about all the many sequential features without being
able to assume such things as determinacy and sequential control, then
the proofs become much more difficult.

One would expect this approach to run into trouble because
critical assumptions underlying the sequential operational
semantics would not hold in the concurrent setting.
For example, on a shared-memory multiprocessor we cannot
assume that (x:=x+1; x:=x+1) has the same effect as
(x:=x+2); and on any real multiprocessor 
we cannot even assume \emph{sequential consistency}---that the
semantics of $n$ threads is some interleaving of the steps
of the individual threads.

We will solve this problem in several stages.
Stage 1 of this plan is the current paper.
Stages 2, 3, and 4 are work in progress; 
the remainder is future work.
\begin{enumerate}
\item We have made the language, the Separation Logic, and our proof
extensible: the set of control-flow statements is fixed (inductive) but
the set of straight-line statements is extensible
by means of a parameterized module in Coq.  We have added to each state
$\sigma$ an \emph{oracle} which predicts the meaning of the extended
instruction (but which does nothing on the core language).  All the proofs we
have described in this paper are on this extensible language.
\item 
\label{contrib1}
We define spawn, lock, and unlock as extended straight-line statements.  We define a 
concurrent small-step semantics that 
assumes noninterference 
(and gets ``stuck'' on interference).
\item From this semantics, we calculate
a single-thread small-step semantics equip\-ped with
the oracle that predicts the effects of synchronizations.
\item We define a Concurrent Separation Logic for \cminor{}
as an extension of the Sequential Separation Logic.
Its soundness proof uses the sequential soundness proof
as a lemma.
\item We will use 
Concurrent Separation Logic to guarantee noninterference
of source programs.  Then
(x:=x+1; x:=x+1) \emph{will} have the same effect as
(x:=x+2).
\item We will prove that the \cminor{} compiler (\compcert{})
compiles each
footprint-safe source thread into an equivalent 
footprint-safe machine-language thread.  Thus, noninterfering source
programs will produce noninterfering machine-language programs.
\item We will demonstrate, with respect to a formal model
of weak-memory-consistency microprocessor, that 
noninterfering machine-language programs give the same results as they
would on a sequentially consistent machine.
\end{enumerate}
\vspace{-13pt}

\section{The Machine-checked Proof}
We have proved in Coq the soundness of Separation Logic for \cminor{}.
Each rule is proved as a lemma; in addition there is a main theorem
that if you prove all your function bodies satisfy their pre/postconditions,
then the program ``call main()'' is safe.
We have informally tested the adequacy of our result by doing tactical proofs
of small programs \cite{appel06:septacs}.  

\begin{tabular}{r@{\quad} p{4.1in}}
Lines & Component\\ \hline
41 & Axioms: dependent unique choice, relational choice,
extensionality \\
8792 & Memory model, floats, 32-bit integers, values, operators, maps 
(exactly as in CompCert \cite{leroy06}) \\
4408 & Sharable permissions, \cminor{} language, operational semantics \\
462 & Separation Logic operators and rules\\
9874 & Soundness proof of Separation Logic
\end{tabular}

\noindent These line counts include some repetition of specifications (between Modules
and Module Types) in Coq's module system.
\vspace{-8pt}

\section{Conclusion}
In this paper, we have defined a formal semantics for the language \cminor. It consists of
a big-step semantics for expressions and a small-step semantics for statements.
The small-step semantics is based on continuations mainly to allow a uniform
representation of statement execution.
The small-step semantics deals with nonlocal control constructs (return, exit)
and is designed to extend to the concurrent setting.

Then, we have defined a Separation Logic for \cminor. It consists of an assertion language
and an axiomatic semantics.
We have extended classical Hoare triples to sextuples in order to take into account
nonlocal control constructs.
From this definition of sextuples, we have proved the rules of axiomatic semantics,
thus proving the soundness of our Separation Logic.

We have also  proved the semantic equivalence between our
small-step semantics and the big-step semantics of the \compcert{} certified compiler,
so the \cminor{} programs that we prove in Separation Logic can be compiled by the
\compcert{} certified compiler.
We plan to connect 
a \cminor{} certified compiler directly to the small-step semantics, instead
of going through the big-step semantics.  

Small-step reasoning is useful for sequential programming languages
that will be extended with concurrent features;
but small-step reasoning about 
nonlocal control constructs
mixed with structured programming (loop) is not trivial.
We have relied on
the determinacy of the small-step relation so that we
can define concepts such as $\absorb{n}{s}{\st}$.

\bibliographystyle{plain}
\bibliography{appel}

\end{document}